# Full Rate Collaborative Diversity Scheme for Multiple Access Fading Channels

Indu Shakya, Falah H. Ali, Elias Stipidis

**Abstract:** User cooperation is a well-known approach to achieve diversity without multiple antennas, however at the cost of inevitable loss of rate mostly due to the need of additional channels for relaying. A new collaborative diversity scheme is proposed here for multiple access fading channels to attain full rate with near maximum diversity. This is achieved by allowing two users and their corresponding relays to transmit/forward data on the same channel by exploiting unique spatial-signatures of their fading channels. The base-station jointly detects the co-channel users' data using maximum-likelihood search algorithm over small set of possible data combinations. Full data rate with significant diversity gain near to two-antenna Alamouti scheme is shown.

**Corresponding Author:**

**Dr. Falah H. Ali,**

**Communication Research Group,**

**School of Engineering and Design,**

**University of Sussex,**

**Brighton, UK**

**BN1 9QT,**

**Tel: +44 (0)1273-678445, Fax: +44 (0)1273-678399**

**Email: f.h.ali@sussex.ac.uk**

# I. INTRODUCTION

User cooperation schemes [1]-[5] that realize spatial diversity by coordinating two or more single antenna terminals, have recently emerged as an attractive alternative to multiple antenna diversity. This is however attained at the loss of rate mostly due to the need of extra channels for relaying. Various schemes have been proposed in the literature to address this problem. For example, a scheme referred to as opportunistic multipath for bandwidth-efficient cooperative multiple access, is proposed in [2] for code division multiple access (CDMA) where each user is assigned two idle relays that forward its estimated data in turn over two consecutive time periods. It exploits the capability of CDMA pseudo noise spreading codes to resolve the multipath from the relays to meet the above objective, however at the cost of increased multiuser interference as the system loading increases. Another approach called superposition coding assisted cooperative multiple access is proposed in [3], where the relaying user superimposes its own data along with source users' data. It assumes full knowledge of channels and their precoding matrix at all user nodes and destination. This requires closed loop operation with feedback channels, leading to a more complex system and some loss in throughput. Another example is space-division relaying (SDR) proposed in [4] that allows two users to exchange their data in first and second periods, and in the third period, space division multiplexing rather than time division is used for simultaneous relaying of users' data. It shows improved rate of $l = 2/3$ compared with $l = 1/2$ e.g. in [1].

In relation to these work, the following contributions are made in this paper: a) a higher rate scheme for cooperative communications is proposed that achieves $l = 1$ with near full diversity, b) collaborative diversity approach is investigated to permit simultaneous transmission of data from more than one user and their corresponding relays using the same channel, c) a low complexity maximum-likelihood (ML) receiver algorithm is used both at relays and the base-station to jointly detect and recover the users' data based on the minimum Euclidian distance criterion over a small number of possible data combinations exploiting the differences in their fading channel gains. To verify the gains, system design for an uplink of orthogonal CDMA is shown, though also applicable to other multiple access schemes e.g. TDMA and O/FDMA.

# II. PROPOSED COLLABORATIVE DIVERSITY SCHEME

To illustrate the principles of our scheme, we consider as an example a group consisting of two collaborating users i.e. $L = 2$ and two corresponding relays as shown in Figure 1 for the k[th] group. Each group is assigned only one orthogonal spreading code $\mathbf{c}_k, 1 \leq k \leq G$. An orthogonal CDMA system can support up to $G$ such groups and each group can hence be independently assigned a single or $L$ users without any mutual interference between groups. The transmission protocol is shown in Table I. In the first period, the users $k1$ and $k2$ transmit their data $b_{k1}$ and $b_{k2}$ using $\mathbf{c}_k$. The data are received at the relays $k1$, $k2$ and the base-station via their corresponding fading channels denoted by $\{g_{k1}^{(1)}, g_{k2}^{(1)}\}, \{g_{k1}^{(2)}, g_{k2}^{(2)}\}$ and $\{g_{k1}, g_{k2}\}$, respectively. The relays perform CDMA despreading followed by joint detection of the co-channel signals using an ML search algorithm based on minimum Euclidian distance criterion over small number of data combinations to give estimates $b'_{k1}$ and $b'_{k2}$. The search complexity for the ML detection is $Q = M^L$, where $M$ is the cardinality of modulation scheme used. In the second period, the relays $k1$ and $k2$ forward the data $b'_{k1}$ and $b'_{k2}$ via $g'_{k1}$ and $g'_{k2}$, respectively to the base-station using the same code $\mathbf{c}_k$. The base-station performs despreading to give $z_k$ and $z'_k$ from the first and second period, respectively. The signals are then diversity combined and jointly detected for co-channel users' data using the ML criterion to give $\hat{b}_{k1}$ and $\hat{b}_{k2}$ with diversity order approaching two. We assume that the users and their relays are located sufficiently far apart from each other so that the channel pairs $\{g_{k1}^{(1)}, g_{k2}^{(1)}\}, \{g_{k1}^{(2)}, g_{k2}^{(2)}\}$ and $\{g_{k1}, g_{k2}\}$ are uncorrelated to ensure that the ML joint detectors are able to distinguish co-channel users' data by exploiting their channel gain differences.

*A. Period 1:- User Data Transmission*

The transmitted signal in this period (shown as solid-line arrows in Figure 1) from the kl[th] user $\mathbf{s}_{kl}$ is given by: $\mathbf{s}_{kl} = \sqrt{P_{kl}} b_{kl} \mathbf{c}_k$, where $P_{kl}$ is the power, $b_{kl}$ is the modulated user data of period $T_b$, $\mathbf{c}_k$ is a unit norm spreading code with chip period $T_c$. The spreading factor is given by $N = T_b / T_c$. The signal received at relay $\mathbf{r}_{kl}, l \in \{1,2\}$, is given by:

$$\mathbf{r}_{kl} = \sum_{l=1}^{L} g_{kl}^{(l)} \mathbf{s}_{kl} + \sum_{u=1, u \neq k}^{G} g_{ul}^{(l)} \mathbf{s}_{ul} + \mathbf{v}, \qquad (1)$$

where $g^{(l)}{}_{kl}$ is flat fading channel gain from the kl$^{th}$ user to it's (kl$^{th}$) relay and $\mathbf{v}$ is the AWGN noise.

*B. Decoding at Relays*

The detection of $b_{kl}$ at the relay (shown as an arrow underneath it in Figure 1) is performed by first CDMA despreading of $\mathbf{r}_{kl}$ with $\mathbf{c}_k$ to give the output signal $z_k$. The interference effect of other groups and their users at the despreader output remains zero due to the use of orthogonal codes between the groups. The data estimate $b'_{kl}$ is then obtained by searching over all $Q$ possible data combinations as follows:

$$b'_{kl} = \arg \min_{b_q^{(l)} \in \{\mathbf{b}_1, \mathbf{b}_2..\mathbf{b}_Q\}} \left| z_k - \sum_{l=1}^{L} b_q^{(l)} g_{kl}^{(l)} \right|^2, \qquad (2)$$

where $Q = 4$ for BPSK data ($M = 2$) and $L = 2$, $\mathbf{b}_q^{(l)} = \{b_q^{(1)},...,b_q^{(L)}\}$ is a possible data combination from the set $\mathbf{b}_q^{(l)} \in [\{-1-1\}, \{-1+1\}, \{+1-1\}, \{+1+1\}]$.

*C. Period 2: Relay Data Forwarding*

In this period, $b'_{kl}$ at each relay is spread using the same $\mathbf{c}_k$ to form transmit signals (shown as dashed-line arrows in Figure 1) given by: $\mathbf{s}'_{kl} = \sqrt{P_{kl}} b'_{kl} \mathbf{c}_k$. The data $b'_{k1}$ and $b'_{k2}$ are forwarded to the base-station using channels $g'_{k1}$ and $g'_{k2}$, respectively and may not always be identical to $b_{k1}$ and $b_{k2}$ due to decoding error at the relays. Although error correction techniques such as CRC codes as in [2] can be used, we do not consider them here for simplicity.

*D. Diversity Combining and Joint Detection at the Base-station*

The received signals at the base-station during the first, $\mathbf{r}$, and the second period, $\mathbf{r}'$, can be written as:

$$\mathbf{r} = \sum_{k=1}^{G} \sum_{l=1}^{L} g_{kl} b_{kl} + \mathbf{v} \quad \text{and} \quad \mathbf{r}' = \sum_{k=1}^{G} \sum_{l=1}^{L} g'_{kl} b'_{kl} + \mathbf{v}. \qquad (3)$$

These signals are despread using the $\mathbf{c}_k$ to obtain output signals $z_k$ and $z_k^{'}$. As can be noted $z_k$ and $z_k^{'}$ consist of two copies of the original data $b_{kl}, l \in \{1,2\}$ via different uplink channels $g_{kl}$ and $g'_{kl}$, respectively. The signals $z_k$ and $z_k^{'}$ are diversity combined and the data are jointly detected. For this purpose, sum of squared Euclidian distance from $z_k$ and $z_k^{'}$ to each possible data combination is calculated to give the final data decision as follows:

$$[\hat{b}_{k1}, \hat{b}_{k2}] = \arg \min_{\mathbf{b}_q^{(l)} \in \{\mathbf{b}_1, \mathbf{b}_2 .. \mathbf{b}_Q\}} \left| z_k - \sum_{l=1}^{L} b_q^{(l)} g_{kl} \right|^2 + \left| z_k^{'} - \sum_{l=1}^{L} b_q^{(l)} g'_{kl} \right|^2 \quad (4)$$

### III. Performance Results

A baseband model of 2-user uplink CDMA $(L=2)$ with BPSK users and binary Walsh Hadamard codes of length $N=16$ are used and the system is assumed fully loaded $(G=N)$. The signal power at user and relay terminals is normalized by $L=2$ so that the new scheme requires the same total power as the existing schemes. We assume all the channels are Rayleigh flat fading and the uplink channels have equal variances i.e. $E\{|g_{kl}|^2\} = E\{|g'_{kl}|^2\} = s^2, \forall k, \forall l$, where $E\{.\}$ is the expectation.

To quantify the gain from cooperation, a ratio of power of a user-to-relay channel $g_{kl}^{(l)}$, compared to that of the user's uplink channel $g_{kl}$, is defined as: $b_k = E\{|g_{kl}^{(l)}|^2\} / E\{|g_{kl}|^2\}$. Another power ratio of a user-to-relay channel $g_{kl}^{(l)}$, compared to that of the co-channel-user-to-relay $g_{ki}^{(l)}, i \neq l$, is defined as: $m_k = E\{|g_{kl}^{(l)}|^2\} / E\{|g_{ki}^{(l)}|^2\}$. It is more likely in practice that the co-channel users are located further apart than they are to the base-station and hence $m_k > b_k$ due to the uniform distribution assumption of users within a cell, however for the purpose of investigating the cooperation gain, we choose $m_k = b_k$ as the worst case scenario of this condition.

Synchronisation is an important practical consideration; therefore two cases are also investigated here: a) perfect synchronisation and b) perfect synchronisation in the first period, but with a timing error in the second period. The timing error signal is modeled as complex Gaussian random variable with standard deviation in fraction of chip period. Channel estimation is also an important practical issue, however it is assumed perfect to investigate the ultimate gain of the proposed scheme irrespective of channel estimation method used, study of which is the subject of future investigations.

*Rate and Diversity Performances:* The efficiency in terms of rate for a cooperative CDMA can be defined as $l = KT_{co} / NT_{nco}$, where $T_{co}$ and $T_{nc}$ are the durations required for the cooperative and the non-cooperative scheme, respectively. The proposed scheme with $G = N$ and $L = 2$, can support $K = LG = 2N$ users, hence, even with $T_{co} = 2T_{nco}$ it achieves full rate $l = 1$ compared with $l = 0.5$ in [1] where $K = N$. Figure 2 shows the BER under perfect synchronisation for different $b_k$ values. BER of existing full rate schemes, single antenna non-cooperative transmission, and that using 2-antenna Alamouti space-time encoding [6], are also shown. It is evident that the proposed scheme achieves significantly improved BER as $b_k$ increases. For example, with $b_k = 10$ dB, $E_b / N_0$ gain of $\approx 7$ dB is achieved at BER of $10^{-3}$. This is attributed to the use of ML joint detectors both at relays and the base-station and spatial diversity of fading channel links. As $b_k$ increases to 30 dB, the BER further improves, approaching within 1 dB of the '2 antenna Alamouti' scheme.

*Effect of Synchronisation Error:* The BER under the synchronisation error at different $E_b / N_0$ values are plotted in Figure 3 for the case of $b_k = 30$ dB. For comparison, BER of 'No Cooperation' scheme for different $E_b / N_0$ values under perfect synchronisation are also plotted. The points where 'No Cooperation' line crosses the proposed scheme at a respective $E_b / N_0$ denote the tolerable timing errors in chips to still retain its performance advantage. For example, with error of as high as 0.25 chips, it can still offer gain.

## IV. Conclusion

A full rate user collaborative diversity scheme with near full diversity is presented. This is shown to be achieved by pairing two users and their corresponding relays to transmit/forward data using the same channel and then jointly detecting the co-channel users' data by exploiting the differences in their channel gains in a collaborative manner. For example, a significant SNR gain of $\approx 9$ dB is shown compared with non-cooperative direct transmission and also within $1$ dB from the two-antenna Alamouti scheme while operating at the full rate of unity. It is also shown that the diversity is retained under modest synchronisation error. Channel estimation technique and fading correlation are important practical considerations to be investigated in future work. Extension to higher group size is also interesting for further investigations.

**Authors' Affiliations:** I. Shakya, F. H. Ali, E. Stipidis (Communications Research Group, School of Engineering and Design, University of Sussex, Brighton, BN1 9QT, UK, Email: f.h.ali@sussex.ac.uk)

**Table Captions:**

Table I: Proposed transmission protocol for the k[th] group with $L = 2$

**Figure Captions:**

Figure 1: System model of the proposed collaborative diversity scheme with $L=2$ for the uplink of CDMA for the k<sup>th</sup> group

Figure 2: BER of the proposed scheme for different $\boldsymbol{b}_k$ values and $\boldsymbol{m}_k = \boldsymbol{b}_k$.

Figure 3: The effect of synchronisation error on the BER of the proposed scheme for $\boldsymbol{b}_k = 30\,\text{dB}$ and $\boldsymbol{m}_k = \boldsymbol{b}_k$

Table I

|  | User k1 | User k2 | Relay k1 | Relay k2 | Base-station |
|---|---|---|---|---|---|
| Period 1 | Transmit $b_{k1}$ | Transmit $b_{k2}$ | Receive $b_{k1}$ | Receive $b_{k2}$ | Receive $b_{k1}+b_{k2}$ |
| Period 2 |  |  | Forward $b'_{k1}$ | Forward $b'_{k2}$ | Receive $b'_{k1}+b'_{k2}$ |

Figure 1:

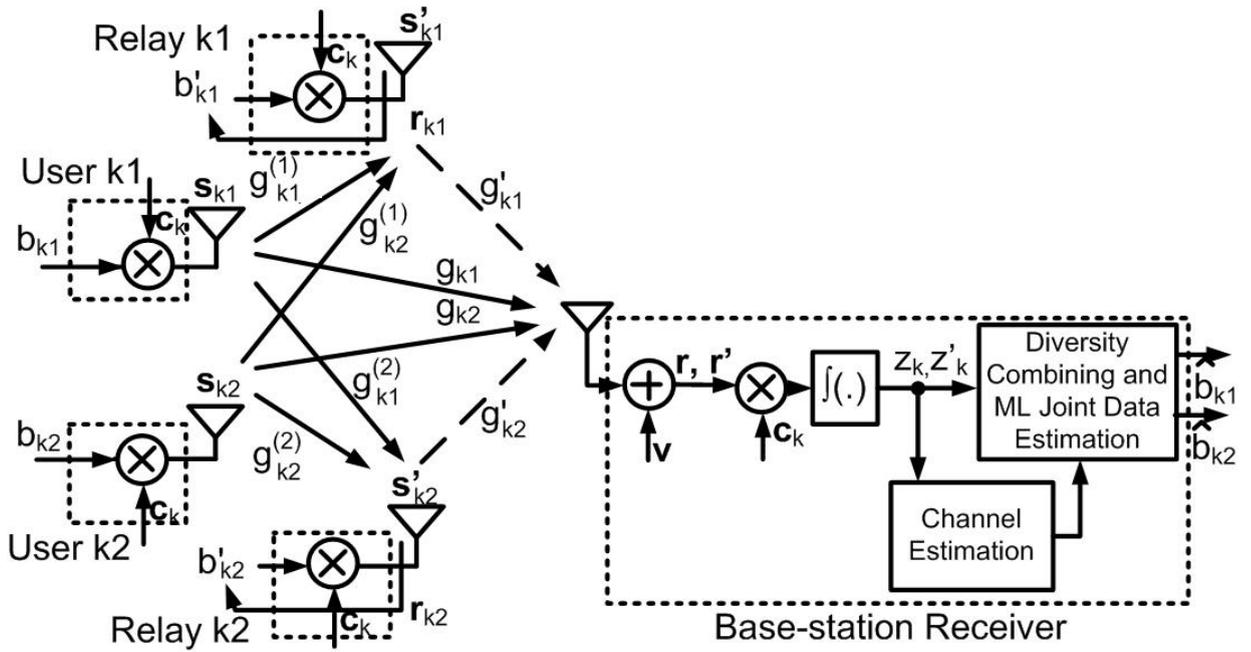

Figure 2:

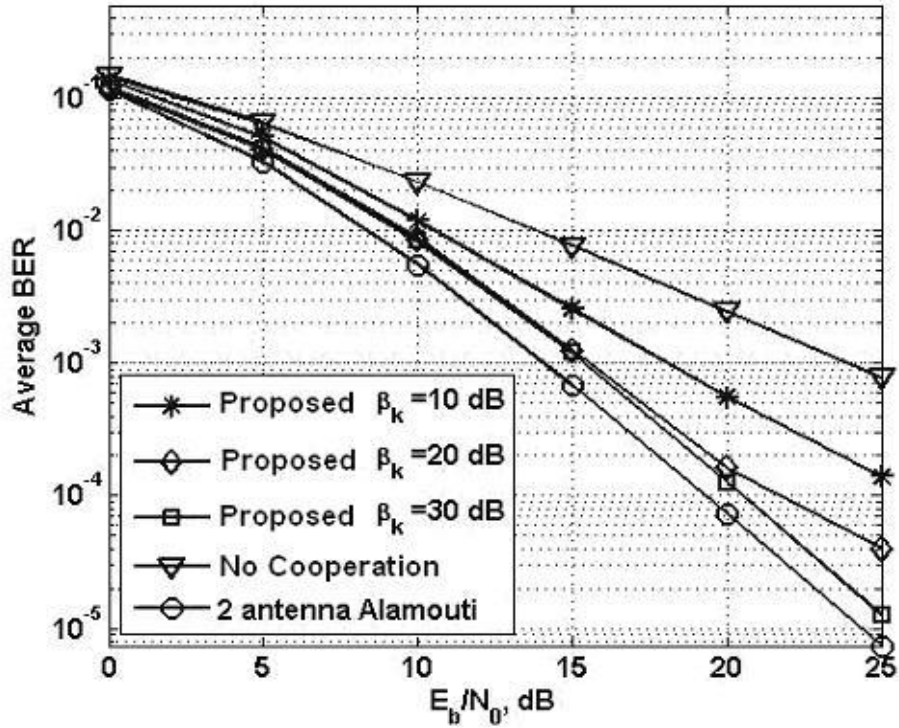

Figure 3:

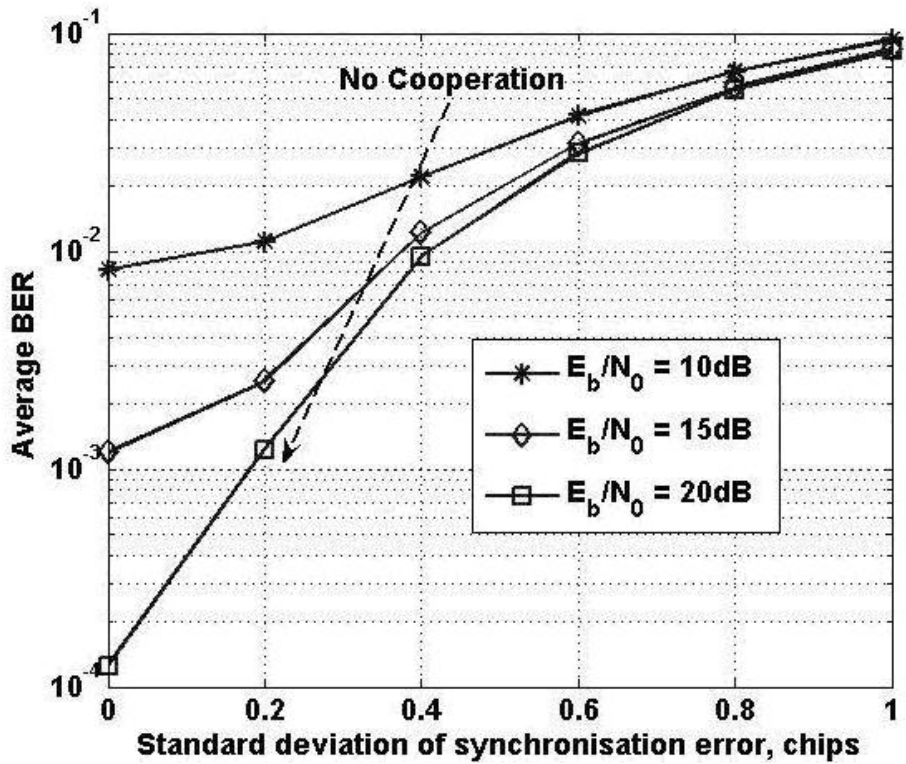